\documentclass[preprint,12pt]{elsarticle}
\usepackage{graphicx,subfigure}
\usepackage[utf8]{inputenc}
\usepackage{amsmath}
\usepackage{amssymb}
\usepackage{units}
\usepackage{booktabs}
\usepackage{array}
\usepackage{multirow}
\usepackage{natbib}
\usepackage{url}
\usepackage{color}

\journal{Journal of Computational Physics}
\begin{document}
\begin{frontmatter}
\title{A fully parallel, high precision, $N$-body code running on hybrid computing platforms}
\author[sap]{R. Capuzzo--Dolcetta}
\ead{roberto.capuzzodolcetta@uniroma1.it} 
\author[sap]{M. Spera}
\ead{mario.spera@uniroma1.it}
\author[sap]{D. Punzo}
\ead{davide.punzo@uniroma1.it}
\address[sap]{Dep. of Physics, Sapienza, University of Roma, P.le A. Moro 1, Roma, Italy}


\begin{abstract}
We present a new implementation of the numerical integration of the classical, gravitational, $N$-body problem based on a high order Hermite’s integration scheme with block time steps, with a direct evaluation of the particle-particle forces. The main innovation of this code (called HiGPUs) is its full parallelization, exploiting both OpenMP and MPI in the use of the multicore Central Processing Units as well as either Compute Unified Device Architecture (CUDA) or OpenCL for the hosted Graphic Processing Units. We tested both performance and accuracy of the code using up to 256 GPUs in the supercomputer IBM iDataPlex DX360M3 Linux Infiniband Cluster provided by the italian supercomputing consortium CINECA, for values of $N \leq 8$ millions. We were able to follow the evolution of a system of 8 million bodies for few crossing times, task previously unreached by direct summation codes.

\end{abstract}

\end{frontmatter}

\section{Introduction}
The study of the motion of $N$ point-like masses having initial positions and velocities $\mathbf{r}_{i0},\mathbf{v}_{i0}, i=1,2,…,N$ interacting through a pair-wise force that depends only on their positions is known as the \textit{N-body problem}. Its applications can be found on both small and large scales starting from nuclear physics up to astrophysical problems. In this latter case, the interaction force is gravity, and, when adopting newtonian gravitational interaction, the problem is referred as the \textit{classical, gravitational, newtonian N-body problem}. Building an appropriate mathematical model for this problem is a crucial task for astrophysicists who want to give an exhaustive representation of objects from planetary systems to galaxy clusters. Actually, an explicit mathematical solution by series, as provided by \citep{wa}, exists (upon some conditions) for an arbitrary $N$ but it requires such an enormous amount of terms to approach convergence that deprives it of any pratical usability. As a matter of fact, the gravitational $N$-body  problem is explicitly solvable only for $N< 3$ while, for $N\geq 3$, the procedure to get a solution is exclusively numerical. 
\par The first numerical simulation of a self gravitating $N$-body system was carried out by \citet{holm}. Accounting for the similar inverse square law scaling with distance of newtonian gravity and the light intensity coming from a light source, he evaluated the gravitational interaction between two model galaxies represented as two groups of 37 lamps. This required entire weeks to accomplish this 74-body simulation. The first numerical simulations of the movement of an $N$-body system were performed by \citet{vho}. Faster numerical integrations were not possible until the years 1960s, when the first digital computers were introduced and Aarseth carried out simulations up to 250 stars \citep{aars}.
From those times, Aarseth has been doing a deep work on high precision simulations of $N$-body systems, producing a series of codes (from NBODY0 up to the last  NBODY7 release \citep{nb7}); a good general reference to direct summation codes is Aarseth's book \citep{grav}.
The well known (at least in the field of astrophysics) GrAvityPipE \textit{GRAPE project} by Sugimoto, Hut and Makino around the end of years 80's (see \url{http://www.astrogrape.org/}) constituted a significant improvement for the $N$-body simulations. The GRAPE boards constitute actual \lq gravity accelerators\rq~ attached to a host workstation and they are still in use (GRAPE-6) allowing high performances for a relatively low cost. Nevertheless, over the last 10 years, the \textit{Graphics Processing Units} (GPUs) are replacing Central Processing Units (CPUs) and GRAPE boards. Actually, the purpose of the GPUs, until a few years ago, was to manage graphics with high levels of data parallelism. Actually, on a screen, many pixels must be updated at the same time and each information is completely independent from each other. A newer GPU concept was born mostly thanks to \textit{CUDA} (Compute Unified Device Architecture), which was developed by Nvidia (2006) and can be considered an extension of some standard high level programming languages like C, C++ and Fortran. It was the begin of a new movement called \textit{General Purpose computing on GPUs} (GPGPUs, see also \url{http://gpgpu.org/}) which is in continuous development and growth because graphic cards are, in general, easy to program and very cheap, exhibiting a huge computational power both in single and double precision precision floating point operations keeping power consumption low. A general review on the hardware and software developments since the 1960s that led to the successful application of Graphics Processing Units (GPUs) for astronomical
simulations is given by \citep{bed12}.
This paper is organized as follows: in Sect. 2 we present the main aspects of the classic, gravitational $N$-body problem; Sect. 3 contains the description of our new $N$-body Hermite's integrator (HiGPUs); in Sect. 4 we illustrate the hardware resources we used for benchmarking the code, whose results are extensively presented and discussed in Sect. 5. Conclusions are drawn in Sect. 6.
In the following we will use the words {\it body}, {\it star} and {\it particle} indifferently because we will consider objects belonging to an $N$-body system always as point-like masses.

\section{The numerical solution of the $N$-body problem}

The numerical solution of the $N$-body problem is a difficult task, because of the so called \textit{double-divergence} of the two-body interaction potential. Actually, the newtonian interaction potential between a point of mass $m_i$ and another of mass $m_j$ is given by 

\begin{equation}
U_{ij}=\frac{Gm_im_j}{\left| \mathbf{r}_j-\mathbf{r}_i \right|} \equiv {\frac{Gm_im_j}{r_{ij}}}=U_{ji},\label{eq:2bpot}
\end{equation}

 where $\mathbf{r}_i$ and $\mathbf{r}_j$ are the position vectors of the \textit{i}-th and the \textit{j}-th star, $G$ is the gravitational constant and $r_{ij}\equiv {\left| \mathbf{r}_j-\mathbf{r}_i \right|}$ represents the euclidean distance between the two particles. As \textit{ultraviolet divergence} we mean the singularity in the $U_{ij}$ potential for very close encounters ($r_{ij}\rightarrow 0$); the \textit{infra-red divergence} corresponds to a never vanishing pair-wise interaction. This double divergence leads to two immediate consequences: 
\begin{enumerate}
\item close encounters ($r_{ij}\rightarrow 0$) yield to an unbound force between interacting stars ($F_{ij}\rightarrow \infty$) producing an unbound error in the relative acceleration;
\item the resultant force acting on every particle of an $N$-body system requires summation over $N-1$ pair-wise contributions, yielding to an $O(N^2)$ computational complexity, which can be overwhelming whenever, as in the relevant astrophysical cases, $N$ is very large (for instance, $N\simeq 10^{11}$ for a typical galaxy).
\end{enumerate}
Moreover, the evaluation of $r_{ij}$ is computationally heavy for it requires the evaluation of the irrational function square root which needs more than one floating point operation (less than ten, with most modern compilators). 
\par The UV divergence is often faced introducing a \textit{softening parameter}, $\epsilon$, in the interaction potential which becomes 
\begin{equation}
U_{ij}=\frac{Gm_im_j}{\sqrt{r_{ij}^2+\epsilon^2}}\label{eq:softpot},
\end{equation} 
that corresponds to substitute point masses with Plummer spheres of scale length $\epsilon$ \citep{Plum11}. In this way, close encounters are smoothed but, of course, this is payed by a loss of resolution at scales of the order of $\epsilon$ and below. On the other hand, to reduce the $O(N^2)$ complexity various strategies are possible (for example mean-field methods). We do not enter here in a discussion on advantages, disadvantages and peculiarities of the many different ways suggested to reduce the computational complexity, we just state what almost unanimously accepted: any of them induce some source of error, which can be systematic and not easy to be controlled. While we address to, e.g., \citep{cd08},\citep{deh11} for a more extended discussion of this topic, and to \citep{hegbook} for a thorough presentation of the field of application of $N$-body simulation 
at scales of $N$ of order one million, we stress here that only the \textit{direct summation} over the whole set of interacting bodies avoids the above mentioned errors but, obviously, it is high computationally demanding.
Clearly, the direct approach to the $N$-body problem is such to take huge advantage by architecture where efficient computing accelerator are coupled to general purpose processors, like in the hybrid CPU+GPU platform. Some authors have already approached this topic, expecially in the frame of the nVIDIA CUDA programming environment; in this context we cite, for instance, \citet{nyland}, \citet{sapp}. In particular, a double parallelization (exploiting both OpenMP over the hosting multicore CPUs and CUDA over the hosted GPUs) has been implemented for a 6th order symplectic $N$-body code by \citet{CDMM11}.

In this paper we choose to couple the power of GPUs as computational accelerators, exploited to evaluate pairwise forces, to the precision guaranteed by a high order (6th) Hermite integrator, as described \citep{nita}, implemented with hierarchically blocked time steps \citep{grav} as a good compromise between accuracy and speed of calculation, as we  show in the following.

\section{The $N$-body code} \label{sec:nbcode}
The main purpose of this paper is to introduce and benchmarking a new fast and friendly $N$-body code suitable for studying the dynamical evolution of stellar systems composed up to 10 millions of stars with the precision guaranteed by direct summation of the pairwise forces. 
\par\noindent
The code, named HiGPUs, is freely available at \url{astrowww.phys.uniroma1.it/dolcetta/HPCcodes/HiGPUs.html}. HiGPUs is, also, part of the AMUSE project (\citep{mcm12}, \url{http://amusecode.org/}). 
HiGPUs is a direct summation $N$-body code mainly aimed at applications in the astrophysical context but easily adaptable to use in different phyisics fields. 
It implements the Hermite's 6th order time integration scheme \citep{nita} with block time steps, allowing both high precision and speed in the study of the dynamical evolution of star systems. The code is written combining tools of C and C++ programming languages and it is parallelized using Message Passing Interface (MPI, \citep{sniretal}), OpenMP \citep{chaetal}, and Compute Unified Device Architecture (CUDA, \citep{san10}, \url{http://developer.nvidia.com/category/zone/cuda-zone}) 
developed by nVIDIA corporation to allow an easy utilization of nVIDIA GPUs. The use of these tools allows our code to exploit in full the most modern hybrid computer platforms which are usually composed by some standard CPUs connected to GPUs acting as high performance computing accelerators.

We have also developed another version of the code based on Open Computing Language (OpenCL, \citep{mun11},  
\url{http://http://developer.amd.com/zones/OpenCLZone/Pages/default.aspx}) which is an open standard for parallel programming of heterogeneous systems,  allowing the use of GPUs of different firms. This latter version has been successfully tested on different AMD GPUs (Radeon series 6xxx and 7xxx) on our private workstations at the Dep. of Physics, Sapienza University of Roma, exhibiting performance comparable or even higher than that obtained with nVIDIA Tesla C2050. 

The coarse-grained parallelization is such that a one-to-one correspondence between MPI processes and computational nodes is established and each MPI process manages all the GPUs available per node. The task decomposition on each node is done taking into account the computational complexity of each section of the Hermite's 6th order scheme, which consists of a \textit{predictor}, an \textit{evaluation}, and a \textit{corrector} step. In the following we will indicate with dots derivatives respect to time. Let us consider a system composed by N stars, and let us assume that the \textit{i}-th particle, has, at time $t_{c,0}$, a position $\mathbf{r}_{i,0}$, a velocity $\mathbf{v}_{i,0}$, an acceleration $\mathbf{a}_{i,0}$, a \textit{jerk} $\dot{\mathbf{a}}_{i,0}$, a \textit{snap} $\ddot{\mathbf{a}}_{i,0}$, a \textit{crackle} $\dddot{\mathbf{a}}_{i,0}$, and an individual time step $\Delta t_{i,0}$. Calling $m$ the number of particles belonging to the same time-block, which have to be evolved to the same time $t_{c,0}+\Delta t_{i,0}$, the generic Hermite's step is composed by various substeps: 
\begin{enumerate}
\item Prediction step, with $O(N)$ complexity: positions, velocities and accelerations of all the stars are predicted using their known values:

\begin{eqnarray}
\mathbf{r}_{i,pred}&=&\mathbf{r}_{i,0}+\mathbf{v}_{i,0}\Delta t_{i,0}+\frac{1}{2}\mathbf{a}_{i,0}\Delta t_{i,0}^{2}+\frac{1}{6}\mathbf{\dot{a}}_{i,0}\Delta t_{i,0}^3+\\
&+&\frac{1}{24}\mathbf{\ddot{a}}_{i,0}\Delta t_{i,0}^4+\frac{1}{120}\mathbf{\dddot{a}}_{i,0}\Delta t_{i,0}^5, \\
 \mathbf{v}_{i,pred}&=&\mathbf{v}_{i,0}+\mathbf{a}_{i,0}\Delta t_{i,0}+\frac{1}{2}\mathbf{\dot{a}}_{i,0}\Delta t_{i,0}^2+\frac{1}{6}\mathbf{\ddot{a}}_{i,0}\Delta t_{i,0}^3+\\
 &+&\frac{1}{24}\mathbf{\dddot{a}}_{i,0}\Delta t_{i,0}^4, \\
  \mathbf{a}_{i,pred}&=&\mathbf{a}_{i,0}+\mathbf{\dot{a}}_{i,0}\Delta t_{i,0}+\frac{1}{2}\mathbf{\ddot{a}}_{i,0}\Delta t_{i,0}^2+\frac{1}{6}\mathbf{\dddot{a}}_{i,0}\Delta t_{i,0}^3.
\end{eqnarray}

\item Evaluation step, with $O(Nm)$ complexity: the accelerations of $m\leq N$ particles as well as their first and second time derivatives are evaluated using the above predicted data. The mutual interaction between the \textit{i}-th particle and the remaining $N-1$ is described by the following relations: \newline 
\begin{eqnarray}
  \mathbf{a}_{i,1}&=&\sum_{\substack{j=1\\j\neq i}}^{N} \mathbf{a}_{ij,1} = \sum_{\substack{j=1\\j\neq i}}^{N} m_{j}\frac{\mathbf{r}_{ij}}{r_{ij}^{3}},\\
  \mathbf{\dot{a}}_{i,1}&=&\sum_{\substack{j=1\\j\neq i}}^{N} \mathbf{\dot{a}}_{ij,1} = \sum_{\substack{j=1\\j\neq i}}^{N}\left(m_{j}\frac{\mathbf{v}_{ij}}{r_{ij}^{3}}-3\alpha_{ij} \mathbf{a}_{ij,1}\right),\\
  \mathbf{\ddot{a}}_{i,1}&=&\sum_{\substack{j=1\\j\neq i}}^{N} \mathbf{\ddot{a}}_{ij,1} = \sum_{\substack{j=1\\j\neq i}}^{N} \left(m_{j}\frac{\mathbf{a}_{ij}}{r_{ij}^{3}}-6\alpha \mathbf{\dot{a}}_{ij,1}-3\beta_{ij} \mathbf{a}_{ij,1}\right), 
\end{eqnarray}

where $\mathbf{r}_{ij}\equiv \mathbf{r}_{j,pred}-\mathbf{r}_{i,pred}$, $\mathbf{v}_{ij}\equiv\mathbf{v}_{j,pred}-\mathbf{v}_{i,pred}$, 
$\mathbf{a}_{ij}\equiv \mathbf{a}_{j,pred}-\mathbf{a}_{i,pred}$, 
$\alpha_{ij} r_{ij}^2 \equiv \mathbf{r}_{ij}\cdot \mathbf{v}_{ij}$, 
$\beta_{ij} r_{ij}^2 \equiv v_{ij}^2+\mathbf{r}_{ij}\cdot \mathbf{a}_{ij} + \alpha_{ij}^{2} r_{ij}^2$

\item Correction step with complexity $O(m)$: positions and velocities of the mentioned $m$ particles to be updated are corrected using the above evaluated accelerations and their time derivatives: 

\begin{eqnarray}
\mathbf{v}_{i,corr}&=&\mathbf{v}_{i,0}+\frac{\Delta t_{i,0}}{2}\left(\mathbf{a}_{i,1}+\mathbf{a}_{i,0}\right)-\frac{\Delta t_{i,0}^{2}}{10}\left(\mathbf{\dot{a}}_{i,1}-\mathbf{\dot{a}}_{i,0}\right)+\\
&+&\frac{\Delta t_{i,0}^{3}}{120}\left(\mathbf{\ddot{a}}_{i,1}+\mathbf{\ddot{a}}_{i,0}\right),\\
\mathbf{r}_{i,corr}&=&\mathbf{r}_{i,0}+\frac{\Delta t_{i,0}}{2}\left(\mathbf{v}_{i,corr}+\mathbf{v}_{i,0}\right)-\frac{\Delta t_{i,0}^{2}}{10}\left (\mathbf{a}_{i,1}-\mathbf{a}_{i,0}\right)+\\
&+&\frac{\Delta t_{i,0}^{3}}{120}\left(\mathbf{\dot{a}}_{i,1}+\mathbf{\dot{a}}_{i,0}\right).
\end{eqnarray}

The individual time steps for $m$ particles are, thus, updated, by mean of the so called \textit{generalized Aarseth criterion} \citep{nita} 

\begin{equation}
\label{indtstp}
\Delta t_{i,1}=\eta\left(\frac{A^{(1)}}{A^{(p-2)}}\right)^{\nicefrac{1}{p-3}},
\end{equation}

where 

\begin{equation}
\label{As}
A^{(s)} \equiv \sqrt{\left|\mathbf{a}^{(s-1)}\right|\left|\mathbf{a}^{(s+1)}\right|+\left|\mathbf{a}^{(s)}\right|^2}.
\end{equation}
In Eqs. \ref{indtstp} and \ref{As}, $p$ is the order of the integration method and $a^{(s)}$ is the \textit{s}-th time derivative of the acceleration. The value of the parameter $\eta$ is linked to the accuracy required for the simulation; we, practically, found that an optimal value for $\eta$ for our Hermite’s 6th order scheme is around 0.6. In any case, the block time steps scheme forces the values of $\Delta t_{i,1}$ to be quantized as powers of two for the sake of an easier control of both the synchronization of all the stars and an efficient code parallelization. In the practical applications of our code, we set as minimum time step $\Delta t_{min} = 2^{-25}\simeq 3.0\cdot 10^{-8}$,  while we fixed as maximum $\Delta t_{max} = 2^{-3} = 0.125$ (both are expressed in program's time unit which coincides with the system crossing time, as defined in Section \ref{sec:result}).

 \end{enumerate} 
We note that the evaluation step is the most expensive in terms of number of operations needed to execute; the predictor step comes after. For this reason we implement both these substeps on the GPUs leaving the less demanding corrector step to the CPUs. In more detail, if $n$ is the number of GPUs used for a simulation, each GPU deals with the predictor of $N/n$ particles and evaluates $3m(N/n)$ accelerations and their first and second order derivatives, collected and reduced from the all set of computational nodes by means of the \verb+MPI_Allreduce()+ functions \citep{sniretal}

The implementation of the evaluation step on a GPU as efficiently as possible requires some consideration on the structure of the GPU. For instance, a GPU nVIDIA Tesla C2050 is capable to run up to $21,504$ threads in parallel, this number corresponding to the combination of 14 Multiprocessors (MPs) which can run up to 1536 resident threads, each. To exploit the full power of such a GPU is, therefore, necessary to run $21,504$ threads in parallel. If the stars to be updated are $m$ and the GPUs to use are $n$, the simplest, but not most efficient, parallelization scheme of the forces calculation would be such to run $m$ threads per GPU and calculate the partial accelerations (and their derivatives) due to $N/n$ bodies. This is simple but if $m$ is small (less than $21,504$) there is not enough work to fully occupy the GPU, causing a significant degrading of performance. In this low $m$ case, in order to increase the number of GPU thread blocks to map to as many MPs as possible, our program tries to reduce the number of threads per block to use in the computation, starting from 128 down, until this number reduces to the minimum possible, set to 32 \citep{nvc}. Anyway, this may be not enough to guarantee a good load of the GPU. To cope with this we introduced a variable, which we call \textit{Bfactor}, acting as factor that multiplies, when necessary, the total number of GPU blocks of threads in order to split further the computation. For example, if $m<21,504$ and $\textit{Bfactor}=B$ we run $m*B$ threads per GPU and the partial accelerations (and derivatives) due to $N/(nB)$ stars are computed. Obviously, before passing the results to the CPU, each GPU has to reduce $B$ blocks of accelerations. This adds a GPU reduction operation to our code but, as we will see, this cost is amortized by what is gained in the evaluation kernel. In any case, the program can recognize the GPU in use, calculate the minimum number of parallel threads to fully load it, and determine the \textit{Bfactor} maximum value by using the formula 

\begin{equation}
B_{MAX}= \left[ \frac{N}{n*TpB}\right]
\end{equation}

where we have used \textit{TpB} to indicate the number of GPU threads per block. The advantages of using this technique will be made clearer in Section \ref{sec:result} where we compare performance (for different values of $m$) of the evaluation step with and without the use of the \textit{Bfactor} variable.

\section{Hardware}
Our new $N$-body code works under standard Linux distributions and requires CUDA (or OpenCL), gcc compiler and OpenMPI implementations. The benchmarking was performed using the IBM iDataPlex DX360M3 Linux Infiniband Cluster available since June 2011 at the Italian supercomputing consortium CINECA (\url{http://www.cineca.it/en}). It consists of 274 computational nodes which exchange data through a Qlogic QDR (40 Gb/s) Infiniband high-performance network. Each node has two GPUs (for a total of 528 nVIDIA Tesla M2070 + 20 nVIDIA Tesla M2070-Quadro), two CPUs Intel Xeon Esa-Core Westmere E5645 running at 2.4 GHz and 48 GB of RAM memory. The operating system is Linux Red Hat EL 5.6 x86\_64 while the version of the gcc compiler installed and tested is the 4.4.4, the CUDA version is the 4.0 and the OpenMPI version is the 1.4.3. We remark that, at present, only about half of PLX is available for scientific users and the maximum number of computational nodes which can be obtained simultaneously is limited to 44 for 6 hours or 22 for 24 hours (\url{https://hpc.cineca.it/content/ibm-plx-user-guide}). This is why some of the benchmarks shown in this work are partially incomplete; nevertheless, upon explicit request, we obtained a few computing hours to run the code on nodes and 128 nodes, in order to test and evaluate performance with 4M and 8M stars. 

\section{Results}\label{sec:result}
In this Section we show the results of our new direct $N$-body code, both in terms of accuracy and performance. We performed a set of N-body simulations with values of $N$ in the range from $32\mathrm{k}$ to
$8\mathrm{M}$ stars, spatially distributed according to the mass density 
\begin{equation}
\rho(r)=\frac{3M}{4\pi b^3}\left(1+\frac{r^2}{b^2}\right)^{-\frac{5}{2}},
\end{equation}  
where $r$ is the distance from the centre and $b$ and $M$ are, respectively, the scale length (also called core radius) and the total mass of the system. The choices $b=1$, $M=1$ and, for the gravitational constant, $G=1$ as physical units for the $N$-body simulations, lead to the so called system crossing time as unit of time for the code, written as
\begin{equation}
t_{c}= \frac{b^{\frac{3}{2}}}{\sqrt{GM}}.
\end{equation}
We used a softening parameter $\epsilon=1.0\cdot 10^{-4}$, which is around 50 times smaller than the minimum closest neighbour average distance for $N=8\mathrm{M}$. 
To perform our benchmarks we chose values of $N$ as powers of two. This is not compulsory but apt to guarantee reaching best optimization. 
 
\subsection{Energy and angular momentum conservation}\label{subsec:cons}

\begin{figure}[!ht]
\centering
{\includegraphics[width=0.8\textwidth]{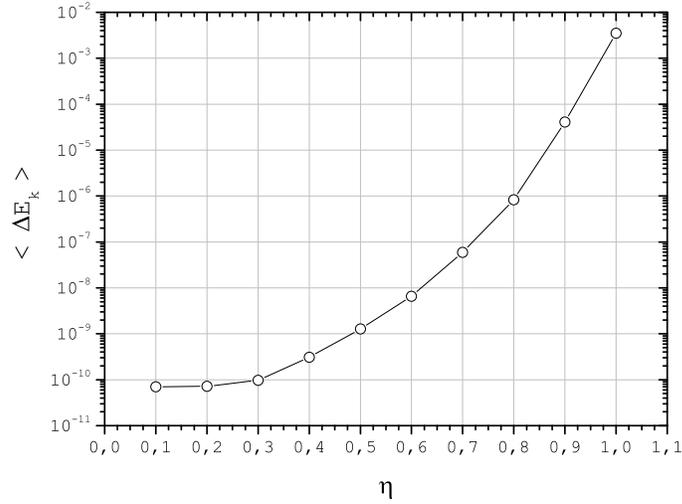}
\includegraphics[width=0.8\textwidth]{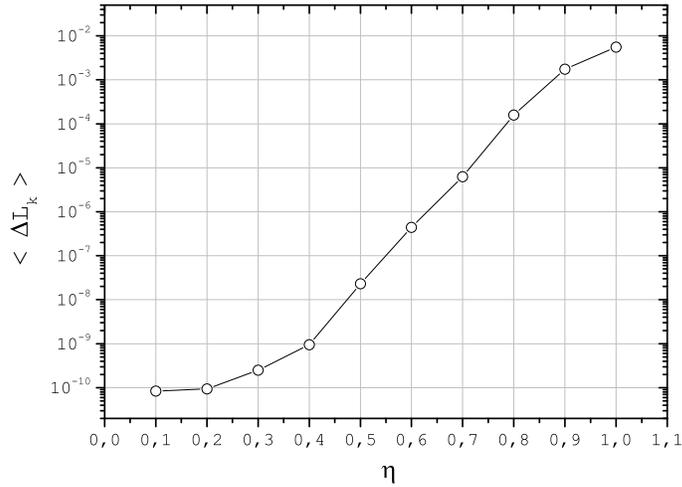}}
\caption{Averaged relative energy (upper panel) and angular momentum (lower panel) errors as a function of the accuracy parameter $\eta$.}
\label{fig:ener}
\end{figure}

The accuracy of our code is controlled by the parameter $\eta$ (see Section \ref{sec:nbcode}). To test the accuracy we run $N$-body simulations with $N=2^k$ with $k$ integer between $\left[15;20\right]$ over 10 time units, checking both the energy and the angular momentum conservation over that time interval. Specifically, the relative errors are calculated using the expressions
\begin{equation}
\Delta E_k = \frac{1}{10}\sum_{i=1}^{10}\left| \frac{E_k(t_{i})-E_k(0)}{E_k(0)}\right|\qquad\qquad    \Delta L_k = \frac{1}{10}\sum_{i=1}^{10}\left| \frac{L_k(t_{i})-L_k(0)}{L_k(0)}\right|,
\end{equation}

where $E_k(t_{i})$ and $L_k(t_{i})$ are, respectively, the total energy and absolute value of angular momentum of the $N=2^k$ system evaluated at  ten times $t_{i}$, which are multiples of the system crossing time. Moreover, the obtained values of $\Delta E_k$ and $\Delta L_k$ are averaged over the five values of $N$. A similar approach to estimate the accuracy of their direct $N$-body code was followed by \citet{ber}. In Fig. \ref{fig:ener} we show the results obtained for different values of $\eta$. As expected the energy error does not depend on $\eta$ when $\eta$ is small enough; for $\eta \lesssim 0.3$  the relative energy error gets an almost constant value around $7.0\cdot 10^{-11}$. Increasing the value of $\eta$ leads to a progressively worse energy conservation, yielding to $\langle \Delta E_k \rangle \simeq 4.0\cdot 10^{-3}$ for $\eta=1.0$. A similar trend is noticed for the angular momentum error. We chose to maintain the energy error for our benchmarks below $5\cdot 10^{-9}$  and the angular momentum error around $5\cdot 10^{-7}$ so we set 
$\eta = 0.6$.

\subsection{Code scalability}\label{subsec:cons}

\begin{figure}[!ht]
\bigskip
\centering
\includegraphics[scale=0.45]{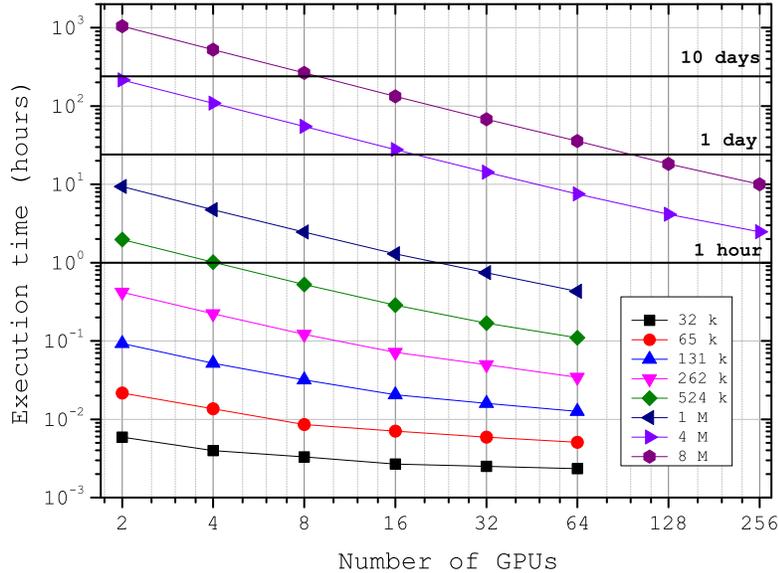}
\caption{Time needed to integrate $N$-body systems ($32k \leq N \leq 8M$) over one time unit using different numbers of nVIDIA Tesla M2070 GPUs .}
\label{fig:times}
\end{figure}

Figure \ref{fig:times} shows the wall clock time needed to integrate an $N$-body system, for different values of $N$, up to one unit of time as a function of the number of GPUs used. 

As relevant result, the total execution time decreases linearly increasing the number of GPUs whenever the number of bodies is large enough (1M, 4M or 8M). The departure from this inverse linear trend is seen for $N\lesssim 262\mathrm{k}$ and when the number GPUs used is greater than 16. This is expected because, when the number of particles per GPU is small ($\lesssim 1000$), the computational load is not enough to exploit the full computational power of the GPUs thus to cover adequately memory latencies, MPI communications, and other non-scalable parts of our code. Another important output of Fig. \ref{fig:times} is that, using 256 GPUs, an integration of a 8M-body system over one time unit is done in less than 10 hours which is, to our knowledge, an unprecedented performance for such kind of high precision, direct summation, $N$-body simulations.

\begin{figure}[!ht]
\centering

\includegraphics[scale=0.45]{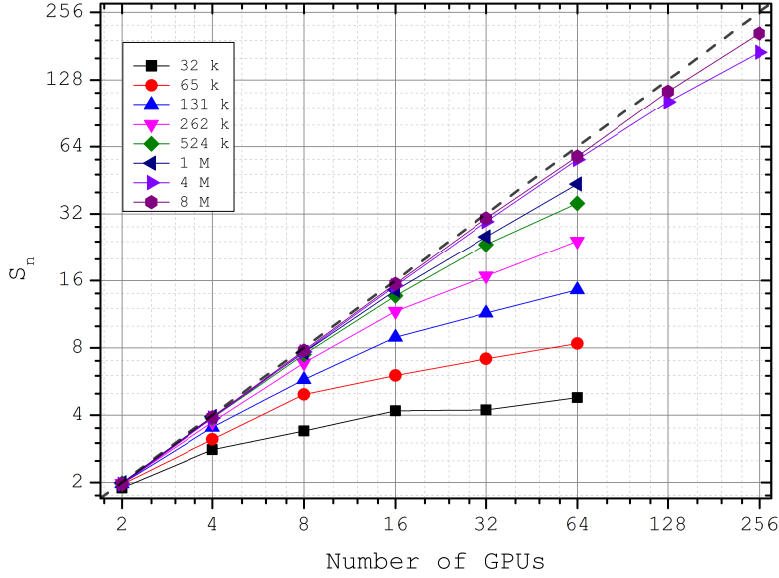}
\caption{The speedup of our code as function of the number of GPUs used. The straight dashed line represents the trend of the perfect speedup.}
\label{fig:effic1}
\end{figure}

\begin{figure}[!ht]
\centering
\includegraphics[scale=0.45]{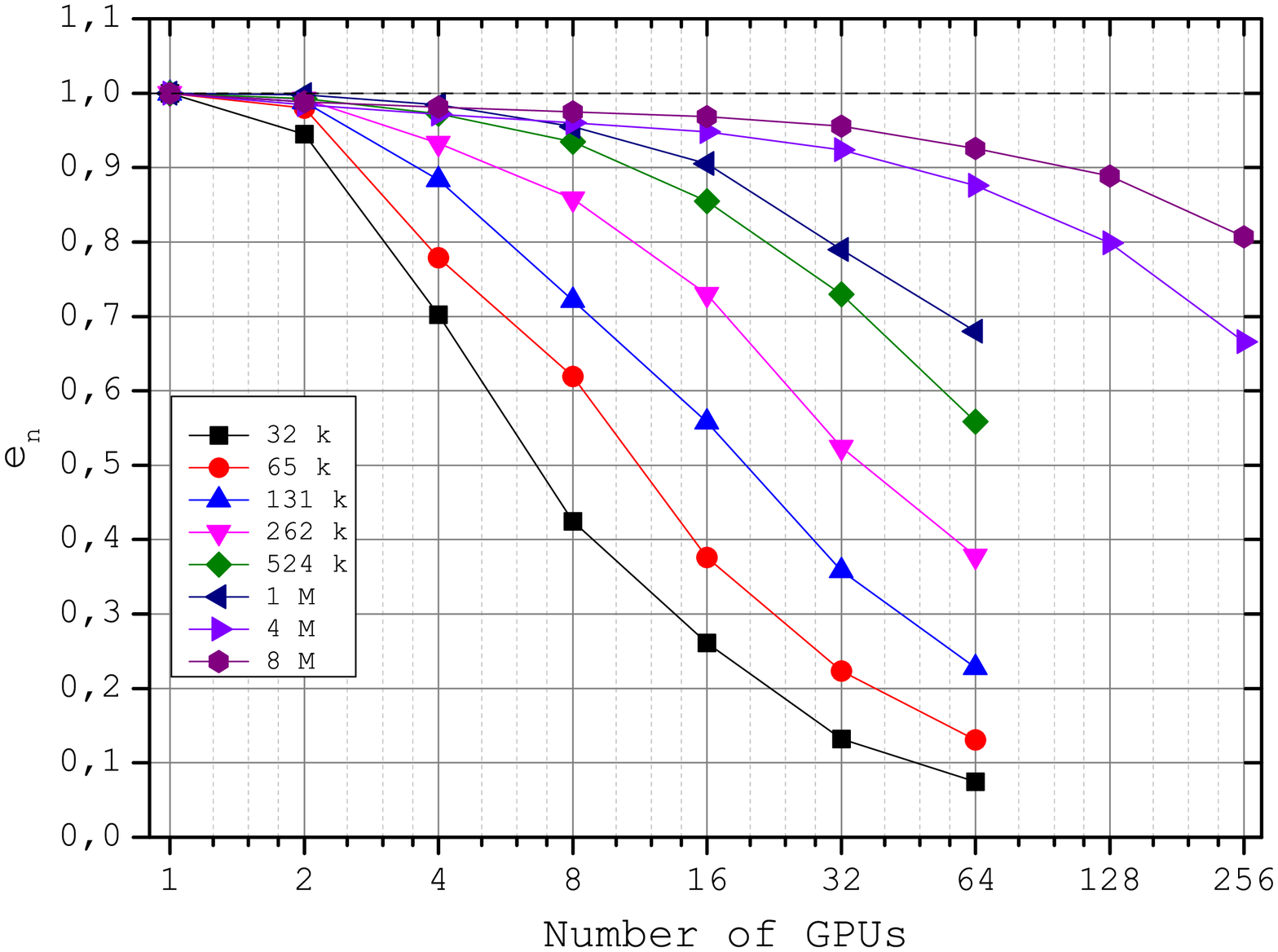}
\caption{The efficiency of our code as function of the number of GPUs used. The horizontal dashed line represents the trend of the perfect efficiency.}
\label{fig:effic2}
\end{figure}

\subsection{Speedup and Efficiency}\label{subsec:speedeff}

A deeper analysis of the performance of our code may be done by mean of the use of parameters like the \textit{speedup}
($S_n$) and the \textit{efficiency} ($E_n$).
The speedup quantifies how faster a parallel algorithm is respect to the corresponding sequential one, and it is defined as: 
\begin{equation}
S_n=\frac{\Delta T_1}{\Delta T_n},\label{eq:speedup}
\end{equation} 
where $\Delta T_n$ is the time spent to execute the program using $n$ computational units (GPUs, in our case). A parallel algorithm is considered to be \lq perfectly\rq$ $   written if the so called \textit{linear speedup} is reached. This ideal situation corresponds to $S_n=n$.

The efficiency $E_n$ indicates how the parallel algorithm exploits the whole available computational resources. It is strongly linked to the speedup and is usually expressed as: 
\begin{equation}
E_n=\frac{S_n}{n}.\label{eq:efficiency}
\end{equation} 
Low efficiencies mean a huge amount of time spent in data communications and/or synchronization events, that are, indeed, real bottlenecks for almost all highly parallel applications.

Figures \ref{fig:effic1} and \ref{fig:effic2} draw how our code is able to integrate up to 8M particles keeping a very good efficiency ($\simeq 0.80$) when using 128 nodes, decreasing to $\sim 0.70$ for $N\simeq 4\mathrm{M}$. Nevertheless, the smaller the number of bodies, the worse the scalability of our code is. This is a behaviour common to this kind of numerical codes, direct consequence of, at least, two different factors:

\begin{enumerate}
\item when the number of particles is small there is not enough work assigned to the generic GPU thread to cover adequately the latencies. This is due i) to the too frequent GPU global memory access compared to the computational load, ii) to the data transfer between GPUs and CPUs, or, in the worst case, iii) to idle GPU cores yielding the performance of the generic GPU to very low levels;
\item increasing the number of GPUs and computational nodes implies the necessity to exchange and reduce data through a network connection, operation which becomes significant in terms of time spent if the number of nodes in use is high (high latencies and low bandwidth whose speed is around 40 Gb/s for the IBM PLX we used).
\end{enumerate}
At the light of these observations, it is clear that the highest efficiency is reached whenever a right balance between the number of GPUs and the size (in terms of $N$) of the astrophysical problem is reached.

\subsection{Code profiling}\label{subsec:optimum}

To obtain a clear picture of what the critical, non scalable, parts of our code are, we divided the operations and tasks  performed in a single time step into 9 parts and we measured their execution times. The schematic representation of our code and its main tasks is given and explained in Table \ref{tab:tab_sections}.

\begin{table}
\begin{tabular}{|c|c|c|c|}
\hline 
Index & Section & Used resource & Notation\tabularnewline
\hline 
\hline 
 & Each node determines the stars &  & \tabularnewline 
 & to be updated and their indexes &  & \tabularnewline
1 & indexes are stored in an array & CPU (OpenMP) & $\Delta t_{next}$\tabularnewline
 & named \emph{next} containing \emph{m} &  & \tabularnewline
 & integer elements &  & \tabularnewline
\hline 
 & Each node copies to its GPUs &  & \tabularnewline 
 & the array containing indexes &  & \tabularnewline
2 & of \emph{m} particles and the & GPU & $\Delta t_{pred}$\tabularnewline
 & predictor step of \emph{N/n} stars &  & \tabularnewline
 & is executed &  & \tabularnewline
\hline 
 & Each node computes the forces &  & \tabularnewline
3 & (and their higher order derivatives) & GPU & $\Delta t_{eval}$\tabularnewline
 & of \emph{m} particles due to \emph{N/n} bodies &  & \tabularnewline
\hline 
 & Each node reduces the calculated &  & \tabularnewline
4 & forces and derivatives & GPU & $\Delta t_{redu}$\tabularnewline
 & of \emph{Bfactor} blocks &  & \tabularnewline
\hline 
5 & Each node adjusts conveniently & GPU & $\Delta t_{repo}$\tabularnewline 
& the reduced values &  & \tabularnewline
\hline 
6 & The CPUs receive the accelerations & GPU $\rightarrow$ CPU & $\Delta t_{DtoH}$\tabularnewline
& from the GPUs &  & \tabularnewline
\hline 
 & The \textrm{MPI\_Allreduce()} functions &  & \tabularnewline
7 & collect and reduce accelerations & CPU(MPI) & $\Delta t_{mpi}$\tabularnewline
 & from all the computational nodes &  & \tabularnewline
\hline 
8 & Corrector step and time step  & CPU & $\Delta t_{corr}$\tabularnewline
 & update for \emph{m} stars &  & \tabularnewline
\hline 
 & The reduced accelerations &  & \tabularnewline
 & (and derivatives) and the corrected &  & \tabularnewline
9 & positions and velocities of \emph{m} & CPU $\rightarrow$ GPU & $\Delta t_{HtoD}$\tabularnewline
 & bodies are passed to the GPUs &  & \tabularnewline
 & of each node &  & \tabularnewline
\hline 
\end{tabular}\caption{The main sections constituing our code, which are performed at each time step. We 
indicate with $n$ the number of GPUs used in the computation. The \textquotedblright{}convenient
adjustment\textquotedblright{} mentioned in the description of the 5th section of our
code refers to the re-organization of the computed and reduced
accelerations and derivatives in one array only (instead of three)
to improve the performance of the subsequent data transfer from the
GPU to the CPU. In this way we execute one bigger copy instead of
three smaller ones.}
\label{tab:tab_sections}
\end{table}

\begin{figure}[!ht]
\bigskip
\centering
\includegraphics[scale=0.45]{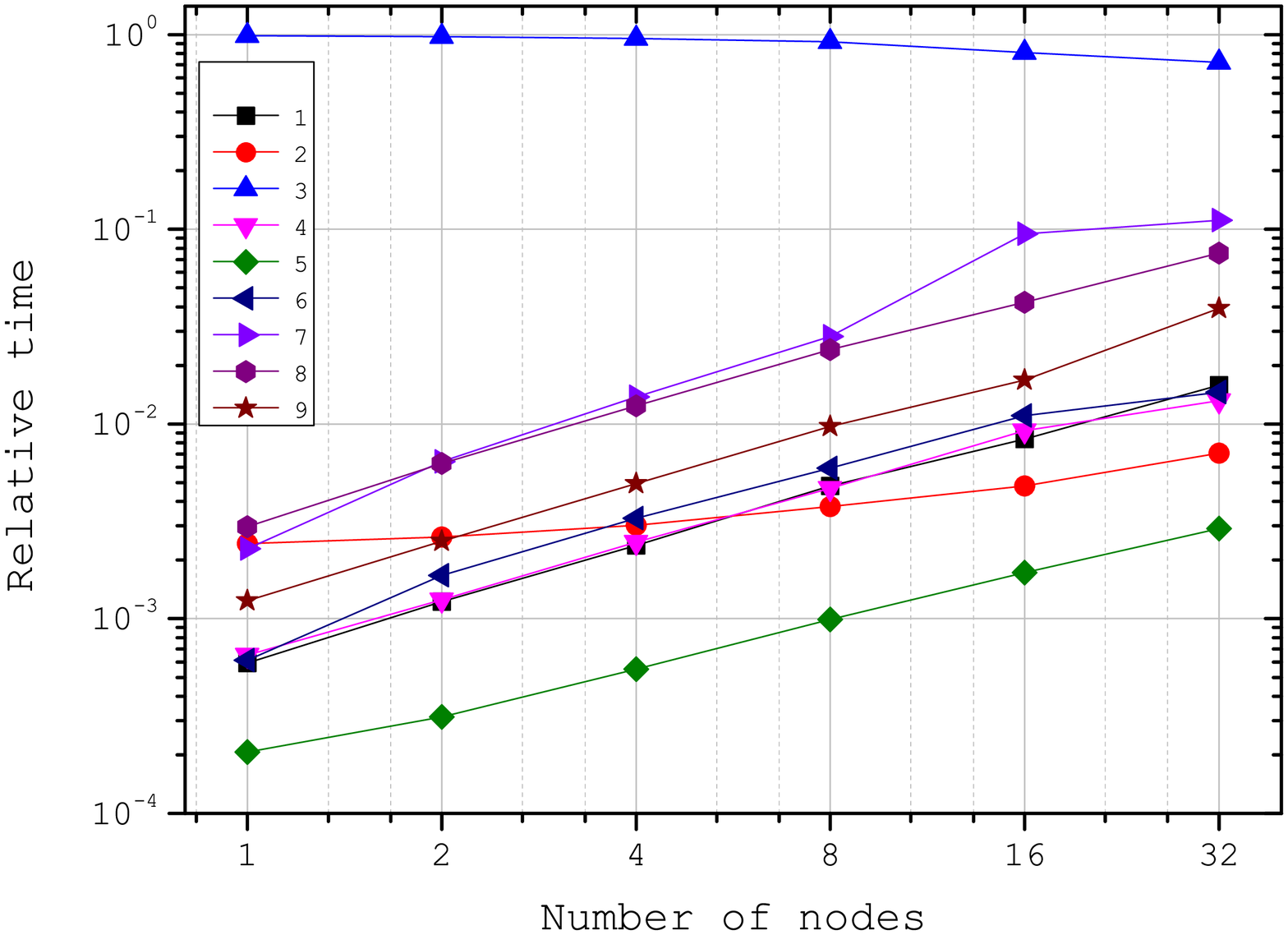}
\caption{Relative (to the total) execution times of the different parts of our code (labeled as in Table \ref{tab:tab_sections}) integrating an $N=1M$ system up to one unit of time using various numbers of computational nodes.}\label{fig:all1M}
\end{figure}

To investigate the performance of the individual sections of our code, we will focus on a system composed of about 1M stars ($N=2^{20}$ to be precise) chosen as reference because it exhibits an average behaviour among all our benchmarks. Following the notation already listed in Table \ref{tab:tab_sections}, we report in Fig. \ref{fig:all1M} the fractional times spent to complete different sections of our code as a function of the number of computational nodes used.

It is worth noted that the force calculation section of the code becomes less important in terms of execution time at increasing the number of nodes, reducing from about 100\%, when using 1 node (2 GPUs) only, to 75\% with 32 computing nodes. Contemporarily, the relative contribution of the other code sections increases, especially the MPI communication part which goes from 0.2\% with 1 node to 10\% when we use 32 nodes. The same happens for the corrector step (7.5\% of the total time with 32 nodes) which, however, has not been yet parallelized and for the CPU--to--GPU data transfer, which contributes for a maximum of 4\% over the total execution time. 

\begin{figure}[!ht]
\bigskip
\centering
\includegraphics[scale=0.45]{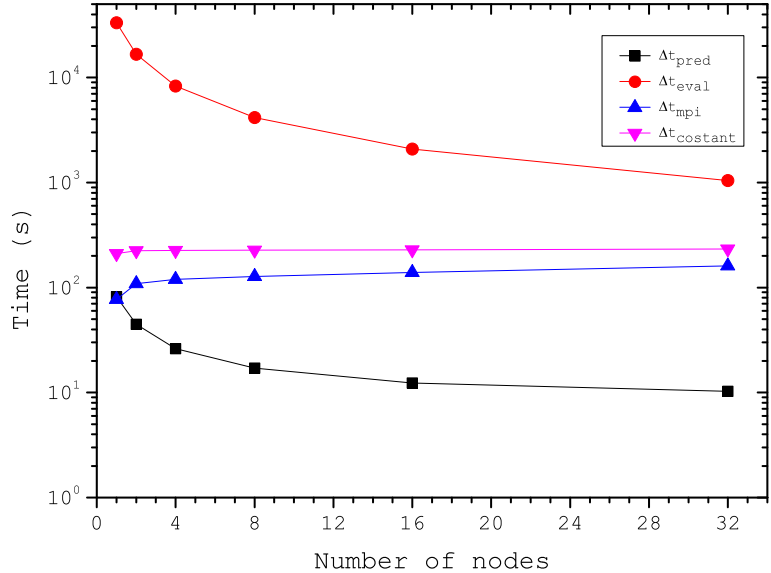}
\caption{The times needed to complete the evaluation step, the predictor step, the MPI communications and the other sections of the code grouped together in the remaining curve. All the times refer to an 1M stars system integrated over one time unit.}\label{fig:sc1M}
\end{figure}

Figure \ref{fig:sc1M} is very helpful to identify possible bottlenecks in the code. It shows the time spent in the various sections of the code (as indicated in Table \ref{tab:tab_sections}) as a function of the number of computational nodes when integrating the 1M-body system over 1 time unit. The figure shows that the two sections that scale increasing the number of GPUs are the evaluation step (which is the most relevant part) and the predictor step.

 The trend of the dependence of the evaluation step on the number of nodes, $n_n$, is well fitted by

\begin{equation}
\Delta t_{n_n,eval}=a\cdot {n_n}^\alpha \label{eq:teval}
\end{equation} with best fitting values $a = 33,213.0\pm1.6\, \mathrm{s}$ and $\alpha = -0.99968\pm0.00030$. This denotes, according to Eq. \ref{eq:speedup}, a very good speedup of our main gpu kernel, at least for $N$=1M and up to 64 GPUs. Moreover, we checked that for $N\in \left[ 32\mathrm{k};8\mathrm{M} \right]$ the value of $\alpha$ remains around $-1$, i.e. we get an approximatively linear speedup for the evaluation step. On the other hand, the $\Delta t_{mpi}$ part of the code, as shown in Fig. \ref{fig:sc1M}, grows with $n_n$ with an almost logarithmic scaling as a result of a complex combination of latency effects, low network bandwidth and inter-node reduction operations. The time spent in this part of the code is fitted by the expression
\begin{equation}
\Delta t_{n_n,MPI}=b+c\log_{10} {n_n}\label{eq:tmpi}
\end{equation} with best parameters $b=85.2\pm5.2\, \mathrm{s}$ and $c=49.3\pm5.7\, \mathrm{s}$. The logarithmic growth of this section is common to all values of $N$ and may reduce significantly the efficiency and scalability of our code when using a large number of GPUs.

At the light of the analysis above it is clear that a faster network connection could improve performance of our code significantly. Moreover, the corrector step (which constitutes about 70\% of the constant part in Fig. \ref{fig:sc1M}) may be furtherly speeded up by its parallelization using, for instance, OpenMP directives.

\subsection{Consequences of block time steps and optimizations}\label{subsec:btso}

\begin{figure}[!ht]
\bigskip
\centering
\includegraphics[scale=0.45]{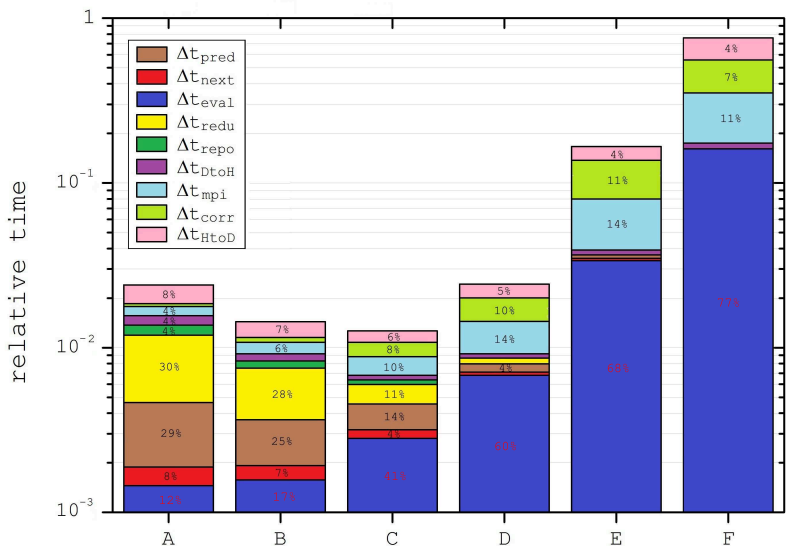}
\caption{Considering the 1M-body reference system, calling $f_m$ the update frequency of $m$ particles over 1 time unit and calling $T_{m,s}$ the time needed to complete the section $s$ of our code for $m$ stars, the percentage value $H_{G}$ of each bar can be obtained as $H_G=\frac{100}{T_{tot}}\sum_s \sum_{m \in G} T_{m,s} \cdot f_m$, where $T_{tot}$ is the total time needed to complete 1 time unit and $G$ indexes the groups $A \: (m \in \left[1;20\right])$, $B \: (m \in \left[21;100\right])$, $C \: (m \in \left[101;1k\right])$, $D \: (m \in \left[1k;10k\right])$, $E \: (m \in \left[10k;100k\right])$, $F \: (m \in \left[100k;1M\right]).$} 
\label{fig:compact}
\end{figure}

Our direct summation $N$-body code is implemented by mean of hierarchically blocked time steps. This implies that the stars to be integrated in a generic time step may vary from 1 to $N$ depending on how the blocks are populated. As a consequence, it is interesting to see what are the group of particles giving the biggest contribution in terms of the total execution time, in order to know where our code needs further optimization. To do this, we measured the update frequencies for the whole set of stars over one time unit (in our 1M-body reference case) using 32 computational nodes and then we multiplied these values by the sum of the times needed to complete each section of the whole time step for those bodies. After grouping particles into 6 groups (labeled with the letters A, B, C, D, E and F) we obtained the results sketched in Fig. \ref{fig:compact}.

Figure \ref{fig:compact} show that, when using 32 nodes for the dynamical integration of 1M stars, the evaluation time can become significantly smaller, for example, than that for the per-block particles determination ($\sim$ 29\% for the A group, brown part) or than that for the GPU reduction of the partial forces due to the presence of the \textit{Bfactor} ($\sim$ 30\% for the A group, in yellow). Moreover, the forces calculation time may be comparable to that needed to complete the predictor step (red) and to that needed to exchange data from the CPU to the GPU (pink). This situation becomes more evident when the number of nodes increases, while it fades, as expected, with larger number of bodies; the two effects tend to compensate, approximately, each other. Therefore, in order to obtain a better performance, it is worth performing a further improvement of our code in the case when the number of stars to be updated is small compared to $N$. 
It must be noted that these results are strongly related to the chosen initial conditions.
Actually, we checked that the inclusion of a massive or super-massive object (black hole) in the N-body system influences the local dynamics such that the contribution of the blocks containing less particles becomes much more significant as previously shown by \citep{mio}.

\begin{figure}[!ht]
\bigskip
\centering
\includegraphics[scale=0.48]{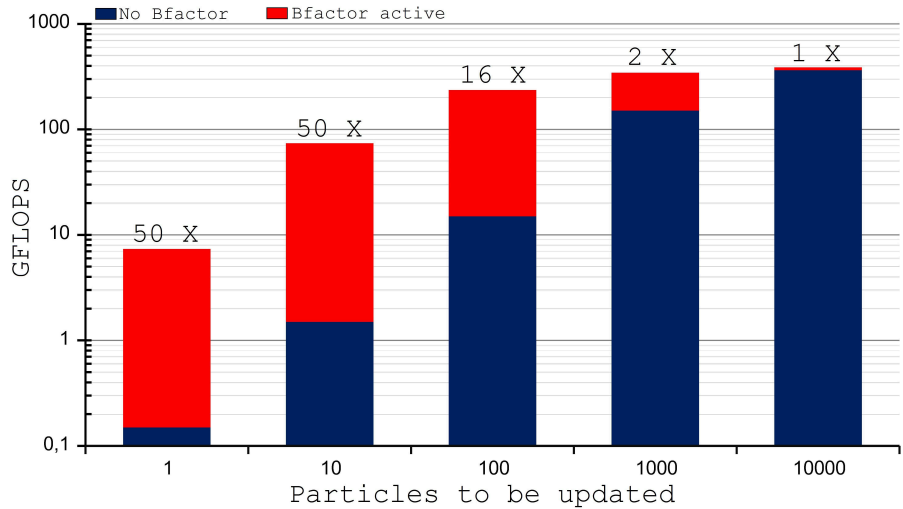}
\caption{Speed (in GFLOPS) achieved updating different numbers of stars using 64 GPUs for a 1M-body system. The red bars indicate the improved performance when the \textit{Bfactor} optimization is active.}\label{fig:bfactor}
\end{figure}

The use of the variable which we indicate as \textit{Bfactor} (see Section \ref{sec:result}) can improve significantly performance in these regimes. In Fig. \ref{fig:bfactor} we show the speed achieved in the forces calculation as a function of the number of bodies to be updated, having 16,384 stars per GPU, in the cases when the \textit{Bfactor} optimization is active and when it is switched off.
This mimics the situation in which an $N$-body system of 1,048,576 stars is run using 64 GPUs (32 IBM-PLX computational nodes). As we see in Fig. \ref{fig:bfactor} the discussed optimization helps to improve performance up to a factor 50 when the number of particles to be updated is less then 20. This means that, referring to Fig. \ref{fig:compact}, the contribution to the total time of the first bar (A) would have been around 50 times larger without introducing a \textit{Bfactor} value becoming the real bottleneck for our applications.

\subsection{GPU memory used by HiGPUs}\label{subsec:gpumem}

\begin{figure}[!ht]
\bigskip
\centering
\includegraphics[scale=0.45]{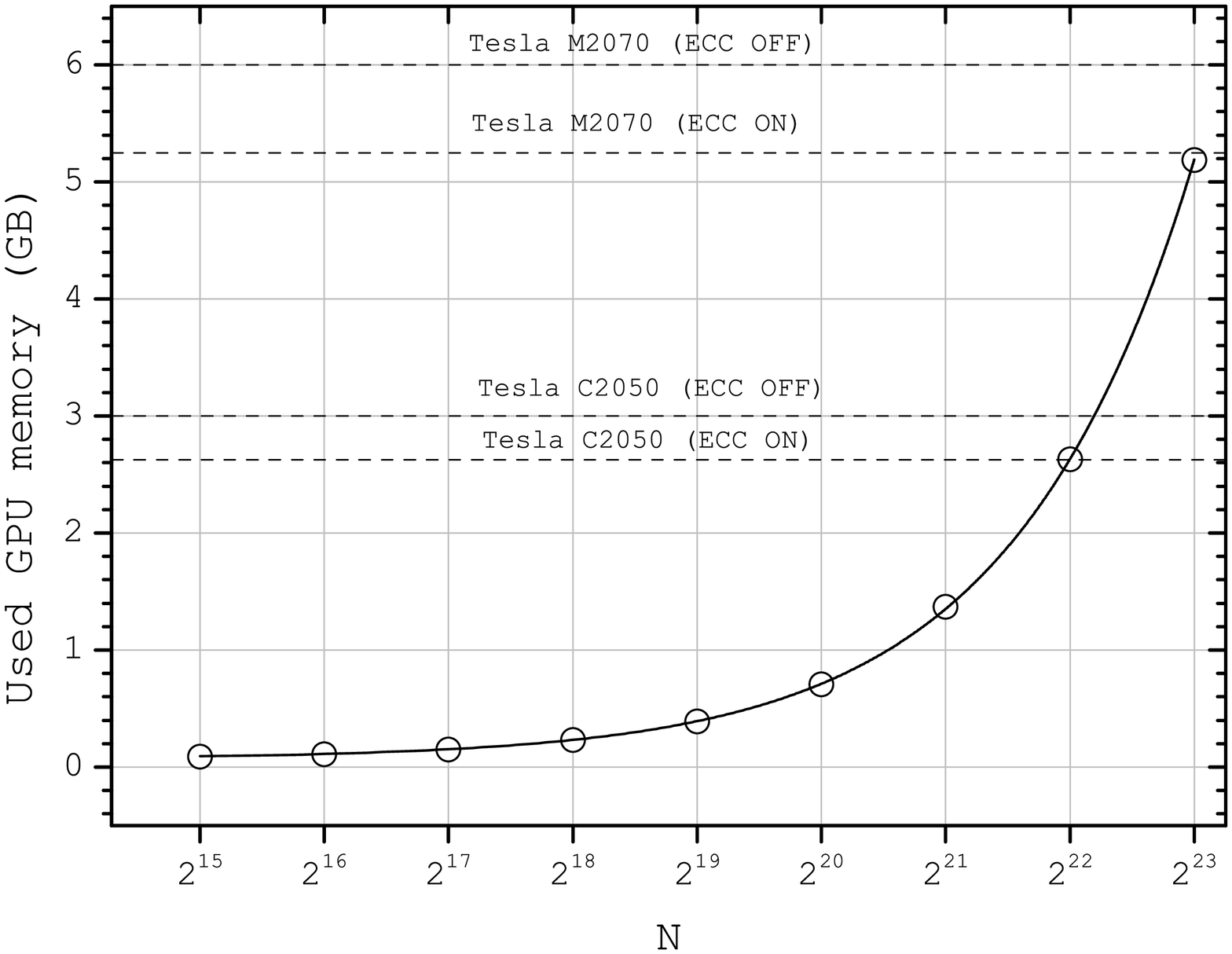}
\caption{Used on-board GPU memory, in GB, as it grows with the number of stars of a generic $N$-body system.}\label{fig:mem}
\end{figure}

As final benchmark, we investigated the maximum GPU memory used as a function of the number of stars to integrate. The results are shown in Fig. \ref{fig:mem}. As we can see, on a GPU Tesla M2070 it is possible to handle up to $N=$ 8M stars, while using a Tesla C2050 $N$ is reduced to 4M.

This is, indeed, a real limit for our code applicability.

\subsection{Hardware maximum performance}\label{subsec:hmp}

The nVIDIA architecture named FERMI organizes the generic GPU as a group of 16 Streaming Multiprocessors of 32 Streaming Processors each for a total of 512 cores which often are called simply \textit{cuda cores}. The GPUs tested in this work (Tesla M2070), using our N-body code, are based on the FERMI architecture having 448 active cuda cores over the 512 potentially available. Each of them has a clock frequency around 1.15 GHz and can execute up to two single precision floating point operations (32 bit) per clock cycle. This means that the theoretical performance peak in single precision can be determined by 
\begin{equation}
S_{peak}=448\, \mathrm{cores}\cdot 2\, \mathrm{flops/cycle} \cdot\, 1.15\, \mathrm{GHz}=1030.4\, \mathrm{GFLOPS}.
\end{equation}
On the other hand, up to 32 double precision floating point operations (64 bit) can be performed per Streaming Multiprocessor, per clock cycle. Having 14 active multiprocessors on a Tesla M2070 we get 
\begin{equation}
D_{peak}=14\, \mathrm{multiprocessors}\cdot 32\, \mathrm{flops/cycle} \cdot\, 1.15\, \mathrm{GHz}=515.2\, \mathrm{GFLOPS},
\end{equation}
 which is exactly half of the single precision peak. It has to be underlined that this compute capability in double precision operations is unprecedented and it constitutes a big jump for high performance computing solutions because, before the advent of the FERMI architecture, the biggest problem to use GPUs for scientific fields was indeed the lack of 64 bit operations. Our GPU main kernel, which calculates the accelerations between stars and their higher order derivatives, uses double precision variables to store predicted positions, accelerations and derivatives, while it computes partial results in single precision mode using 32 bit registers. We prefer to use this hybrid strategy although other authors (see for example \citep{sapp} or \url{http://crd-legacy.lbl.gov/~dhbailey/mpdist/}) sometimes emulate double precision operations storing one 64 bit number in two 32 bit numbers yielding to 14 the number of significant digits and letting the use of GPUs which do not natively support double precision arithmetics (rare nowadays). To estimate how much our code is capable to exploit the FERMI architecture we measured its peak performance. To do this, we counted how many floating point operations are enclosed in our evaluation kernel (the most expensive section in terms of computational load) and then we divided this value by the time needed to execute it, obtaining the performance expressed in GFLOPS. Other authors use different strategies to count operations \citep{elsen,nyland} but we prefer to refer to Table \ref{tab:flops}.

\begin{table}[h]
\begin{center}
\begin{tabular}{|c||c||c|}
\hline
\textbf{Operation} & \textbf{CUDA expression} & \textbf{Equivalent flops}\\
\hline
$a\pm b$ & $\mathrm{a\pm b}$ & 1 \\
$a\cdot b$ & $\mathrm{a*b}$ & 1 \\
$\frac{1}{\sqrt{a}}$ & $\mathrm{rsqrt(a)}$ & 4 \\ 
$\frac{a}{b}$ & $\mathrm{a/b}$ & 5 \\
$a^b$ & pow(a,b) & 9 \\
\hline
\end{tabular}
\caption{The number of floating point operations required by the operations most relevant for our code.}\label{tab:flops}
\end{center}
\end{table}
Table \ref{tab:flops} has been built following the Table 5-1 of the CUDA C programming guide \citep{nvc} together with the information given by the whitepaper of the FERMI architecture \citep{white}. 
Specifically, we can notice that the power elevation operation is very expensive, and it must be avoided, as much as possible, because it is implemented using a combination of one base-2 logarithm, one base-2 power elevation and one multiplication. Following Table \ref{tab:flops} we counted 15 double precision operations plus 82 single precision operations in our main kernel. Therefore, we estimate the theoretical peak achievable, per GPU, by the formula 
\begin{equation}
R_{peak}= \frac{82\cdot S_{peak}+15\cdot D_{peak}}{82+15} \simeq 950\,\mathrm{GFLOPS}.
\end{equation}
This is of course a pure ideal value because it does not consider any kind of memory latency, communication and/or read and write operations which, in general, can reduce performance significantly. The formula that we derived to count GFLOPS in our main kernel is the following 
\begin{equation}
R\simeq\frac{97\cdot N \cdot m}{10^9\cdot \Delta T_{ker}(N,m)}\mathrm{(GFLOPS)} 
\end{equation}
where $N$ is the total number of stars that form our N-body system, $m$ is the number of particles to be updated and $\Delta T_{ker}(N,m)$ is the kernel execution time. A similar formula has been used by other authors \citep{nita,spur}. 
We reached a performance over 100 TFLOPS using 256 Tesla M2070 with $N=2^{23}\simeq 8.4\times 10^6$ stars, which  corresponds to $\sim 400$ GFLOPS per GPU, that is around 40\% of the claimed peak GPU performance. This is a very good result for this kind of astrophysical computations, especially in this case of heavy use of the 
double precision arithmetics, at least comparable to what obtained by other authors with similar (in structure) $N$-body codes (see \citep{spur}).

\section{Conclusions}
Composite architectures based on computational nodes hosting multicore Cnetral Processing Units connected to one or more Graphic Processing Units have been shown by various authors to be a clever and efficient solution to supercomputing needings in different scientific frameworks. Actually, these architectures are characterized by a high ratio between performance and both installation cost and power consumption. A practical proof of this is that some of the most powerful systems in the TOP 500 list of world's supercomputer are based on that scheme. They are indeed a valid alternative to massively parallel multicore systems, where the final computational power comes by the use of a very large number of CPUs, although each of them has a relatively low clock frequency.
It is quite obvious that a full exploit of the best performance of the CPU+GPU platforms requires codes that clearly enucleate a heavy computational kernel, to be assigned in parallel to the GPUs acting as \lq number crunchers\rq~ which release, periodically, their results to the hosts.
In physics, the study of the evolution of systems of objects interacting via a potential depending on their mutual distance falls into this category. 

In this paper we presented and discussed a new, high precision, code apt to simulating the time evolution of systems of $N$ point masses interacting with the classical, pair, newtonian force. The high precision comes from both the evaluation by direct summation of the pairwise force among the system bodies and by a proper treatment of the multiple space and time scales of the system, which means resorting to an individudal time-stepping procedure and resynchronizations, as well as using a high order (6th) time integrator. 

In this paper we discussed the implementation of our fully parallel version of a direct summation algorithm whose O($N^2$) computational complexity is dealt with by GPUs acting as computational accelerators in the hosting nodes where multicore CPUs are governed and linked via MPI directives. The code, called HermiteIntegratorGPUs (HiGPUs, available to the scientific community on \url{astrowww.phys.uniroma1.it/dolcetta/HPCcodes/HiGPUs.html} or in the frame of the AMUSE project on \url{amusecode.org}) shows a very good performance both in term of scaling and efficiency in a good compromise between precision (as measured by energy and angular momentum conservations) and speed.
We performed an extensive set of test simulation as benchmarks of our code using the PLX composite cluster of the CINECA italian supercomputing inter-university consortium. We find that the integration of an $N = 8,000,000$ bodies is done with an 80\% efficiency, that is a deviation of just 20\% from the linear speedup when using 256 nVIDIA Tesla M2070. This corresponds to less than 10 hours of wall clock time to follow the evolution of the 8M body system up to one internal crossing time, performance, at our knowledge, never reached for such kind of simulations. This means that with HiGPUs it is possible to follow the evolution of a realistic model of Globular Cluster (a spherical stellar system orbiting our and other galaxies and composed by about 1,000,000 stars packed in a sphere of about 10 pc radius) with a 1:1 correspondence between number of real stars in the system and simulating particles over a length of about 10 orbital periods around the galactic center in few days of simulation. 
These kind of simulations will allow, for instance, a thorough investigation of open astrophysical questions that may involve, in their answer, the role of globular clusters and globular cluster systems in galaxies. We cite the open problem of the origin of Nuclear Clusters as observed in various galaxies, like our Milky Way. Some authors (e.g., \citep{Mil04} and \citep{Bek07}) suggested a dissipational, gaseous origin while others (\citep{CD93}, \citep{CDV97}) indicate as more realistic a dissipationless origin by orbital decay and merging of globular clusters, as already numerically tested in \citep{CDM08} and \citep{Aetal12}).

One limit in the use of our code is the GPU memory: with a 6GB RAM, as in the case of nVIDIA Tesla M2070, the upper limit in $N$ is $\sim 8,400,000$, which is, anyway, a number sufficiently large to guarantee excellent resolution in the simulation of most of the astrophysically interesting cases involving stellar systems. 
A the light of the results obtained in this paper, we are convinced that it is worth testing some other commercially available high-end GPUs apt to work in a scientific environment, like, for instance, the AMD of the HD6970 and 7970 series, which we showed (\url{astrowww.phys.uniroma1.it/dolcetta/HPCcodes/HiGPUs.html}) to be absolutely competitive in terms of performance/cost ratio. This suggests as clever the idea of adopting for department sized high end computing platforms the solution of few tens of nodes composed by exacore CPUs connected to 2 up to 4 HD7970 GPUs acting as computing accelerators; all this at a very reasonable cost, both on the purchase cost and power consumption.  
\section{Acknowledgments}
We acknowledge the class C grant number HP10COVQZA provided by the italian consortium for supercomputing (CINECA, Casalecchio, Italy) which allowed us to perform the simulation tests presented here.
\bibliographystyle{elsarticle-num-names}
\bibliography{capuzzodolcetta}

\end{document}